# Detection of Clones in Digital Images

**Minati Mishra**
Department of Information and Communication Technology
Fakir Mohan University, Balasore, Odisha, India

**Flt. Lt. Dr. M. C. Adhikary**
Department of Applied Physics and Ballistics
Fakir Mohan University, Balasore, Odisha, India

**ABSTRACT**

During the recent years, tampering of digital images has become a general habit among people and professionals. As a result, establishment of image authenticity has become a key issue in fields those make use of digital images. Authentication of an image involves separation of original camera outputs from their tampered or Stego counterparts. Digital image cloning being a popular type of image tampering, in this paper we have experimentally analyzed seven different algorithms of cloning detection such as the simple overlapped block matching with lexicographic sorting (SOBMwLS) algorithm, block matching with discrete cosine transformation, principal component analysis, discrete wavelet transformation and singular value decomposition performed on the blocks (DCT, DWT, PCA, SVD), two combination models where, DCT and DWT are combined with singular value decomposition (DCTSVD and DWTSVD. A comparative study of all these techniques with respect to their time complexities and robustness of detection against various post processing operations such as cropping, brightness and contrast adjustments are presented in the paper.

## 1. INTRODUCTION

Photographs were considered to be the most powerful and trustworthy media of expression and were accepted as proves of evidences in a number of fields such as forensic investigations, investigation of insurance claims, scientific research and publications, crime detection and legal proceedings etc. But with the availability of easy to use and cheap image editing software, photo manipulations became a common practice. Now it has become almost impossible to distinguish between a genuine camera output and a tampered version of it and as a result of this, photographs have almost lost their reliability and place as proves of evidences in all fields. This is why digital image tamper detection has emerged as an important research area to separate the tampered digital photographs from their genuine counterparts and to establish the authenticity of this popular media [1].

Images are manipulated for a number of reasons and all manipulations may not be called tampering or forging. According to Oxford dictionary, the literary meaning of 'tampering' is interfering with something so as to make unauthorized alterations or damages to it [2]. Therefore, when images are manipulated to fake a fact and mislead a viewer to misbelieve the truth behind a scene by hiding an important component of it or by adding new components to it, it is called a tampering; not the simple manipulations involving enhancements of contrast, color or brightness.

### 1.1 Active Vs Passive Detection Techniques

Active tampering detection techniques such as semi-fragile and robust watermarking techniques require some predefined signature or watermark to be embedded at the time of image creation whereas, the passive methods neither require any prior information about the image nor necessitate the pre embedding of any watermark or digital signature into the image. Hence the passive techniques are more preferred over the active methods. Though a carefully performed tampering does not leave any visual clue of alteration; it is bound to alter the statistical properties of the image and the passive tamper detection techniques try to detect digital tampering in the absence the original photograph as well as without any pre inserted watermark just by studying the statistical variations of the images [3].

*1.1.1 Passive-Blind Detection Techniques*

Passive detection again can be guided or blind depending upon whether the original copy of the image is available for comparison or not. Most of the time, it has been seen that once an image is manipulated to fake some fact, the original image is generally deleted to destroy the evidence. In situations where neither the original image is available nor the image was created with a watermark embedded to it; tamper detection and image authentication becomes a challenging problem. In such cases, passive-blind tamper detection methods can be used to detect possible tampering. In this paper we concentrate on passive-blind methods of cloning detection. The rest of the paper is organized as follows:

Different types of tampering methods are discussed in section 2; different techniques of cloning detection are discussed in section 3, performance evaluation and experimental results are given in section 4 and finally a summary of the experimental studies are presented in section 5.

### 2. Types of Tampering

Based on whether the manipulation is performed to the visible surface of the image or to invisible planes, the manipulation techniques can be classified broadly classified into two types: tampering and Steganography. Again, based on whether the tampering is performed by making changes to the context of the scene elements or without the change of the context, tampering can be classified as context based and content based tampering. In the second case, the recipient is duped to believe that the objects in an image are something else from what they really are but the image itself is not altered [4].

The context based image tampering is generally achieved by copy-pasting scene elements of an image into itself or to other and hence called the copy-move forgery. If an image tampering is performed by copy-pasting a part of an image to itself so as to conceal some object or recreate more instances of the objects in the scene then the process is called cloning. On the other hand if the forged image is created by copy-pasting a part of one image into another then the process is known as splicing.

## 2.1 Image Splicing

In image splicing, a part of an image copied and pasted onto another image without performing any post-processing smoothing operation. By Image tampering, it generally means splicing followed by the post-processing operations so as to make the manipulation imperceptible to human vision. The image given in Figure.1 is an example of image splicing. The image shown in the newspaper cutout is a composite of three different photographs given at the bottom. The White House image is rescaled and blurred to create an illusion of an out-of-focus background on which images of Bill Clinton and Saddam Hussein are pasted [4, 5].

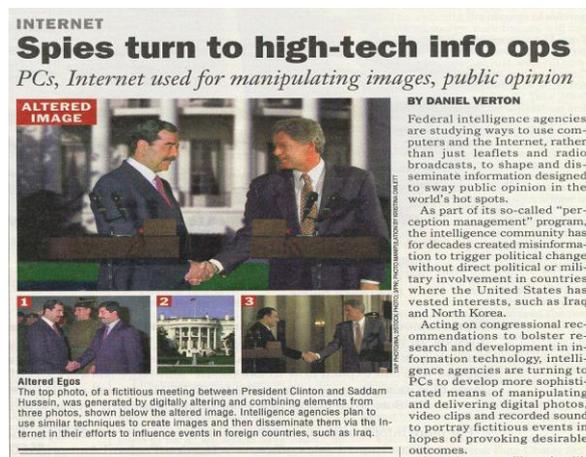

**Figure.1: Spliced image of Bill Clinton with Saddam Hussein**

Because the stitched parts of spliced images come from different images those might have been be taken in different lighting conditions and backgrounds and might have gone through transformation processes such as zooming, cropping, rotation, contrast stretching so as to fit to the target image therefore, careful study of the lighting conditions and other statistical properties can reveal the tampering.

International Journal of Computer Science and Business Informatics

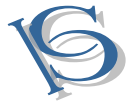

IJCSBI.ORG

## 2.2 Cloning

Cloning or copy-move forgery is a type of image tampering where a part of the image is copy-pasted onto some other part of the same image generally to hide some objects in the scene or to recreate few more instances of some specific objects in an image [3]. It is one of the most commonly used image manipulation techniques. The image in Figure.2 (a) is a clone of the image Figure.2 (b). The person on the scene is hidden carefully copy- pasting and blending a part of the scenery. Similarly, image given in Figure.2 (c) is a clone of Figure.2 (d) where another instance of the gate is recreated copy-pasting a part of the original image.

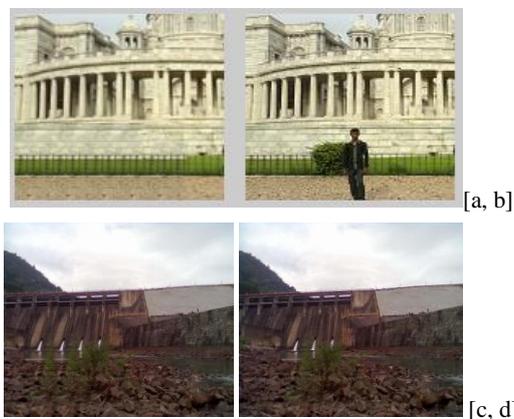

[a, b]

[c, d]

**Figure.2: Images on the left are clones of the right side images**

When done with care, it becomes almost impossible to detect the clone visually and since the cloned region can be of any shape and size and can be located anywhere in the image, it is not computationally possible to make an exhaustive search of all sizes to all possible image locations. Hence clone detection remains a challenging problem in image authentication.

## 3. Techniques of Clone Detection

## 3.1 Exhaustive Search Method

Given an image, the task here is to determine if it contains duplicated regions of unknown location and shape. In an exhaustive search approach, it is required to compare every possible pairs of regions with each other to locate duplicate regions, if any. Though this is the simplest approach for detecting clones in a digital image, the computational time is very high so as to be effective for large size images [5].

## 3.2 Block Matching Procedures

### 3.2.1 Overlapped Block Matching

In this method, the test image of size *(M* x *N)* is first segmented into *(M-b+1)* x *(N-b+1)* overlapping blocks by sliding a window size (*b* x *b*) along the image from top-left corner to right and down by one pixel [6]. Then the blocks are compared for matches. Figure.3 shows the result of this method with a block size of 8x8 pixels. In image given in Figure.3b, the regions marked in red indicate the copy-pasted regions whereas in Figure 3.e the regions given in orange are copied into regions shown in bluish green. Figure.3d is created making multiple copies of a part of the image given in Figure.3f and then cropping the copied regions so as to create a smooth, visually non-detectable forgery. The result therefore, consists of fewer orange blocks in comparison to the number of green blocks. Though this method successfully detects the tampered regions, as can be seen from the results, gives some false positive cases (the region in the sky). The false positives are generated as natural images sometimes have regions with similar pixel intensities. Other problems associated with this method are: (1) dealing with time required to compare large number of blocks. Though, this method requires less number of steps to detect the clones in comparison to the exhaustive search still, the time complexity remains as large as $O\ (b^2 R^2)$, where, $R=(M-b+1)$ x $(N-b+1)$ is the number of blocks and $b^2$ is the size of each block. For example, an image of 128x128 pixels can produce as many as 14641, 15129, 15625 and 15876 blocks of size 8x8, 6x6, 4x4 and 3x3 respectively and direct comparison of each block with each other will require lots of computation time.

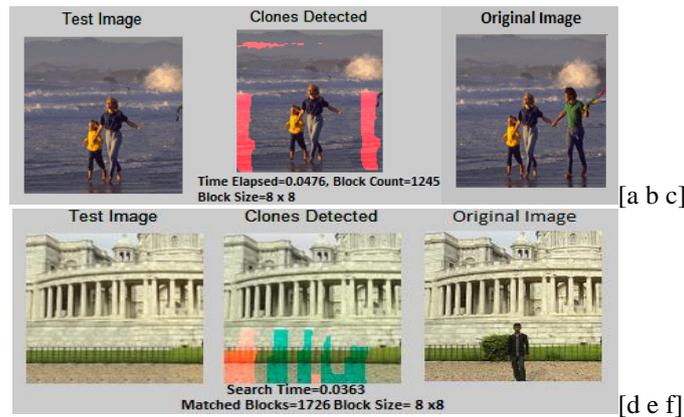

[a b c]

[d e f]

**Figure.3: [a, d] Cloned images, [b, e] duplicate regions detected, [c, f] Original Images**

The second problem is: what should be the optimal block size? The experiments to detect clone blocks in images are performed with multiple block sizes and results are shown in the following Figure.4. It is clear from the experimental results that smaller the block sizes, more better the detection of duplicate regions. But if the block size becomes very small then some false matches are also obtained as in case of the false matches detected (magenta dots and blocks in the lower grass area and in the white sky areas) in the following figure for block size of 3x3, 4x4. Therefore, a good clone detection algorithm should be able to detect a duplicate region even if it is of very small size and at the same time should minimize both the number of false positives as well as computation time. It has been seen that selection of an appropriate block size can help recognizing smaller duplicate regions and by careful design of the block matching step and dimension reduction, the computational efficiency of the algorithm can be improved.

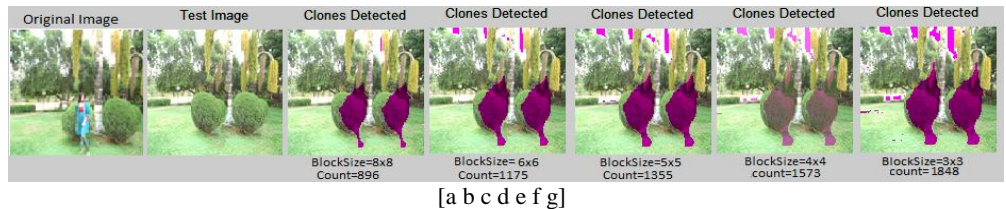

[a b c d e f g]

**Figure.4: Overlapped block matching performed with multiple block sizes**

*3.2.2 Elimination of False Positives by Measuring Block Shift distances*

The false positives can be eliminated by considering image blocks that are at a constant distance, instead of looking for whole duplicated regions as all the blocks of two duplicate regions are likely to be shifted by a fixed distance. Therefore, the tampering decision can be made calculating the shift distances for all matched blocks and then seeing if there are more than a certain number of similar image blocks within the same distance. For example, in the following Figure.5(b) and Figure.5(c), the wrong matches, as detected in the sky area of Figure.5(a) and Figure.4(g), are successfully eliminated by considering the number of blocks shifted through a fixed distance and comparing against the threshold frequency (TH >= 100, in this case).

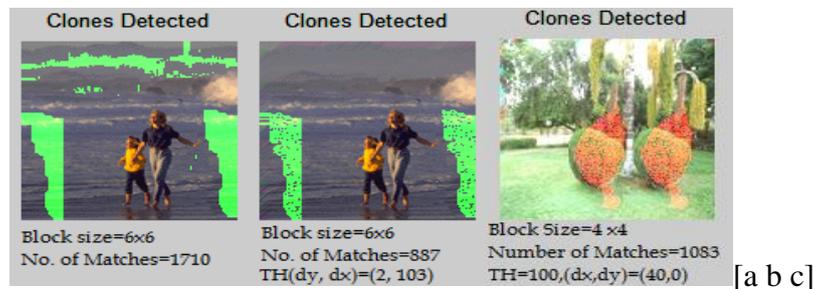

[a b c]

**Figure.5: Elimination of False Positives measuring the Block Shifts**

The measures of various block shifts along x-axis (dx) and y-axis (dy) with the number of blocks shifted (frequency) along each direction for images given in Figure.5 (b) and Figure.5(c) are given in table.1 (a) and table.1 (b) below. It can be seen from the first table that that 94 blocks are shifted just by a single unit along the x-axis and 10 blocks are shifted by 4 units along x-axis and 1 unit along y-axis. Similarly, in the 2$^{nd}$ table, 51 blocks are shifted by 1 pixel along x-direction. All these duplicate blocks represent similar blocks in a natural image, not clones and hence are discarded.

**Table 1: Frequency of block shifts along a distance (dx, dy )**

| | | | | | | | | | | | | | | | | | | | | | | | | | | | | | | | | | | | | | |
|---|---|---|---|---|---|---|---|---|---|---|---|---|---|---|---|---|---|---|---|---|---|---|---|---|---|---|---|---|---|---|---|---|---|---|---|---|---|
| Freq: | 94 | 7 | 7 | 1 | 2 | 2 | 1 | 1 | 1 | 2 | 2 | 1 | 6 | 1 | 1 | 1 | 1 | 1 | 1 | 1 | 1 | 2 | 4 | 10 | 4 | 1 | 2 | 3 | 4 | 1 | 2 | 4 | 1 | 3 | 1 | 1 | 1 3 1 1 1 1 |
| dy: | 0 | 0 | 0 | 0 | 0 | 0 | 0 | 0 | 0 | 0 | 0 | 0 | 0 | 0 | 0 | 0 | 0 | 0 | 0 | 0 | 0 | 0 | 1 | 1 | 1 | 1 | 1 | 1 | 1 | 1 | 1 | 1 | 1 | 1 | 1 | 1 | 1 | 1 1 1 1 1 |
| dx: | 1 | 2 | 3 | 5 | 7 | 9 | 11 | 12 | 13 | 14 | 15 | 16 | 17 | 20 | 27 | 28 | 29 | 35 | 39 | 64 | 107 | 2 | 3 | 4 | 6 | 9 | 10 | 13 | 15 | 21 | 23 | 24 | 25 | 27 | 28 | 29 | 41 | 43 48 50 54 55 |
| Freq: | 2 | 2 | 2 | 2 | 1 | 2 | 2 | 1 | 3 | 3 | 2 | 2 | 1 | 1 | 1 | 1 | 1 | 1 | 127 | 1 | 1 | 2 | 1 | 1 | 2 | 1 | 2 | 1 | 1 | 7 | 1 | 2 | 1 | 1 | 1 | 2 | 3 | 1 1 1 1 1 |
| dy: | 1 | 1 | 1 | 1 | 1 | 1 | 1 | 1 | 2 | 2 | 2 | 2 | 2 | 2 | 2 | 2 | 2 | 2 | 2 | 3 | 3 | 3 | 3 | 3 | 3 | 3 | 4 | 4 | 4 | 4 | 4 | 4 | 5 | 5 | 5 | 5 | 8 | 11 14 14 15 18 |
| dx: | 57 | 58 | 63 | 72 | 74 | 88 | 101 | 109 | 7 | 9 | 12 | 15 | 34 | 35 | 38 | 46 | 74 | 96 | 103 | 6 | 13 | 48 | 50 | 80 | 84 | 94 | 22 | 37 | 46 | 54 | 76 | 105 | 2 | 11 | 25 | 100 | 3 | 8 64 74 3 20 |

[a]

| | | | | | | | | | | | | | | | | | | |
|---|---|---|---|---|---|---|---|---|---|---|---|---|---|---|---|---|---|---|
| Freq: | 51 | 1 | 1 | 1 | 2 | 2 | 127 | 2 | 1 | 3 | 1 | 1 | 1 | 1 | 1 | 2 | 1 | 1 |
| dy: | 0 | 0 | 0 | 0 | 0 | 0 | 0 | 1 | 1 | 1 | 1 | 1 | 1 | 7 | 13 | 14 | 19 | 20 | 40 |
| dx: | 1 | 8 | 9 | 10 | 13 | 23 | 40 | 0 | 1 | 3 | 4 | 59 | 71 | 45 | 13 | 33 | 17 | 15 | 57 |

[b]

*3.2.3 Improving the Search Time through Vectorization and Lexicographic Sorting*

The search time can be highly reduced by representing each block as a vector or a row of a matrix *A*. As there are *(M-b+1)* x *(N-b+1)* number of overlapped blocks of size *b* x *b* in an image of size *M* x *N* therefore, *A* will have *R= (M-b+1)* x *(N-b+1)* rows of *l= $b^2$* elements each. Now by sorting the rows of the matrix *A* in lexicographic order, the similar blocks can be arranged into successive rows of the matrix and can be easily identified with minimum comparison steps without required to compare each row with each other row of the matrix. The lexicographic ordering will require *O ($lRlog_2R$)* steps in case of merger sort or *O (lR)* steps in case of bucket sort is used for the purpose. Many authors represent the time complexity of lexicographic ordering as *O ($Rlog_2R$)* by considering *l* negligible in comparison to *R*. But, when the block size increases the value of *l* increases, requiring more computational steps. In our experiments, we found that the computation time is greater for block sizes greater than 8x8 in comparison to those less than it.

**3.3 Dimension Reduction through DWT**

The decomposition of images using basis functions that are localized in spatial position, orientation, and scale (e.g., wavelets) have proven extremely useful in image compression, image coding, noise removal, and texture synthesis [7]. Therefore, by first decomposing the image into wavelets by DWT and then considering only the low frequency (LL) component of the transformed coefficients which will contain most of the image information, the number of rows of the matrix can be further reduced [8]. This reduces the size of the image to *M/2* x *N/2* pixels and hence the number of rows of the matrix *A* to one-fourth [9]. The following Figure.6 shows the block diagram of a three-level DWT decomposition of an image and Figure.7 shows the steps of the DWT based method.

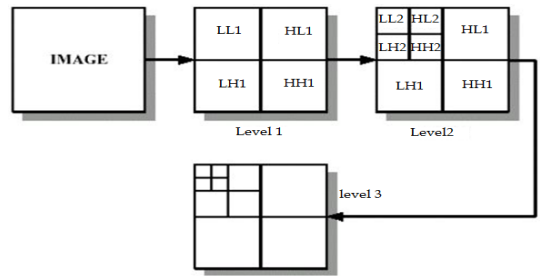

**Figure.6: Decomposition of an Image through DWT**

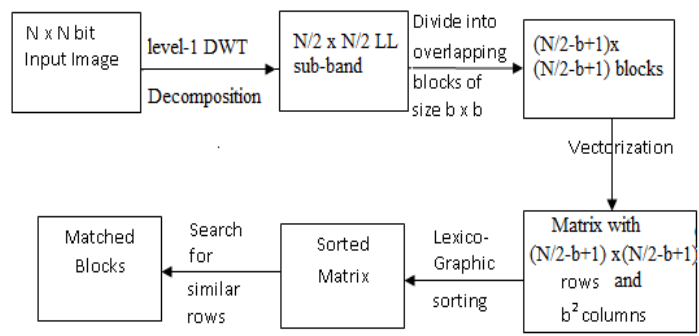

**Figure.7: Block Diagram of Clone Detection through DWT**

*3.3.1 Further Reduction in feature Dimension through SVD*

Singular value decomposition (SVD) is a method for transforming correlated variables into a set of uncorrelated ones that better expose the various relationships among the original data items. At the same time, it is a method for identifying and ordering the dimensions along which data points exhibit the most variation. Once it is identified where the most variation is, it is possible to find the best approximation of the original data points using fewer dimensions. SVD is a method for data reduction where a rectangular matrix $B_{mn}$ is expressed as the product of three matrices - an orthogonal matrix $U$, a diagonal matrix $S$, and the transpose of an orthogonal matrix $V$ as follows[10]:

$$B_{mn} = U_{mm} S_{mn} V^T_{nn} \qquad (1)$$

Where, $U^T U = I$, $V^T V = I$; the columns of $U$ are orthonormal eigenvectors of $BB^T$, the columns of $V$ are orthonormal eigenvectors of $B^T B$, and $S$ is a diagonal matrix containing the square roots of eigenvalues from $U$ or $V$ in descending order [10].

After reducing the total number of vectors (rows) of $A$ to $1/4^{th}$ through DWT, the feature dimension of the matrix (the number of columns) can be reduced from $b^2$ to $b$ by decomposing each block through SVD and considering only the diagonal *b* elements of *S* for comparison in the matching step. Therefore, the matrix *A* now can be viewed as a matrix with *R/4* rows and *b* columns requiring much less search time in comparison to the original matrix. SVD can also be combined with DCT for robust and efficient detection.

*3.3.2 Robust Detection through DCT and PCA*

The overlapped block matching method succeeds only when the duplicate blocks have similar gray values (color intensities) but fails if the pixel intensities of the copied region differ from the original region due to contrast and brightness adjustments as in case of Figure.11 (a) where a part of the image (from bottom right corner is copied and pasted into the bottom left by reducing the pixel intensities. The block matching procedure fails because in this case the source and target regions though have similar values but no more have same values for the pixel intensities. The source (region) pixels values vary from the target pixels with some constant. To detect the matched blocks in such cases, the matching step can be performed after DCT or PCA applied to blocks [5, 6]. Figure.8 shows the block diagram of the DCT based algorithm.

The DCT coefficients *F (u, v)* of a given image block f(x, y) of size *N x N*, can be calculated using the formula

$$F(u,v) = \sum_{x=0}^{N-1}\sum_{y=0}^{N-1} f(x,y)\alpha(u)\alpha(v)\cos\left[\frac{(2x+1)u\pi}{2N}\right]\cos\left[\frac{(2y+1)v\pi}{2N}\right] \quad (2)$$

Where, 
$$\alpha(k) = \begin{cases} \sqrt{\frac{1}{N}} & \text{if } k = 0 \\ \sqrt{\frac{2}{N}} & \text{if } k = 1,2.. \ N-1 \end{cases}$$

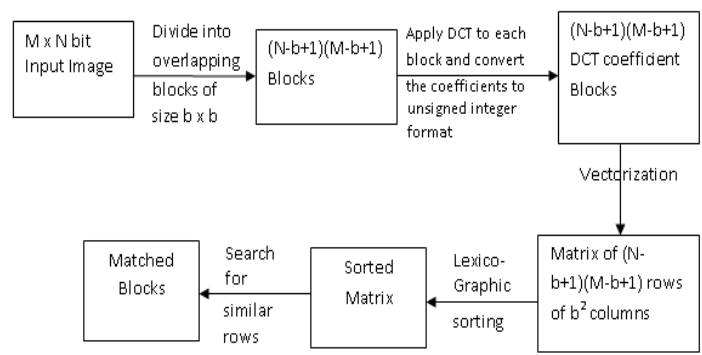

**Figure.8: Steps of DCT based Robust Detection Method**

After DCT is performed to the blocks, $1/4^{th}$ of the low frequency components of each block can be considered for comparison discarding the rest $3/4^{th}$ elements. By this way the size of each block reduces to $b^2/4$ and at the same time the process becomes robust against intensity changes. The blocks, in step3, can also be represented alternatively with a fewer elements by performing principal component analysis (PCA) to each block. PCA is an orthogonal linear transformation that uses orthogonal transformation to convert a set of observations of correlated variables into a set of values of linearly uncorrelated variables called principal components [11]. By considering first few principal components of the data, the size of each block reduces to b and this makes the detection process robust against intensity changes, as well.

## 4. EXPERIMENTAL RESULTS AND DISCUSSIONS

To conduct the experiments, a number of cloned images are created by copy-pasting, cropping, blending parts of some test images. Figure.9 gives results of our experiments with their search times. All the test images considered for this study are square images and preferably fall into three sizes; 128 x 128, 256 x 256 and 512 x 512 pixels. Most of the test images are either grayscale images or converted to gray scale using the formula:

$$Gray = 0.2126R + 0.7152G + 0.0722B \qquad (3)$$

| Original Image | Test Image | Clones Detected (Block size=4x4) | | | | | |
| --- | --- | --- | --- | --- | --- | --- | --- |
| | | SimpleOBM | SVD | DCT | DWT | DCTSVD | DWTSVD |
| | clone1.bmp | Time=.0472 count =1027 | Time=.0368 Count=1162 | Time=.0394 count= 1085 | Time=.0320 count=129 | Time=.0341 count=1197 | Time=.0279 Count= 112 |
| | clone2.bmp | Time=.1312 count=1752 | Time=.0460 count=1754 | Time=.0488 Count=1798 | Time=.0337 count=317 | Time=.0365 count=1753 | Time= .0325 count= 317 |
| | clone3.bmp | Time=.1243 count=1573 | Time=.0447 count=1574 | Time=.0942 count=1625 | Time=.0321 count=226 | Time= .0435 count=1601 | Time= .0313 count= 226 |
| | C11.bmp | Time=.0459 count=1071 | Time=.0406 count=1041 | Time=.0425 count=1373 | Time=.0318 count=199 | Time=.0401 count=1074 | Time=.0305 count=149 |

**Figure.9: Detection of Clones in Different Images using Different Methods**

International Journal of Computer Science and Business Informatics

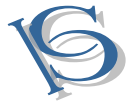

IJCSBI.ORG

In some cases, the R, G and B channels are processed separately for finding the matched regions and then the results are combined afterward. The experiments are performed on more than a hundred of color and gray scale test images of 128x128 pixels or 256 x 256 pixels sizes and it is found from the results obtained that the DWT based search method optimizes the search time as well as effectively locates the duplicate regions. DWT followed by SVD further reduces the search time while preserving the accuracy. Of course, for the DWT based methods to be effective, the size of the cloned region should be at least four times the block sizes e.g, for a block size of 4 x 4 pixels, the clones should be a minimum of 8 x 8 pixels or more else the method fails to detect any tampering. A comparison of computation times for the four test images of figure.9 is shown in a bar chart in Figure.10. The horizontal axis of the chart denotes the image numbers and the vertical axis represents the search times. The individual bars represent the search time taken by a particular image with respect to a selected algorithm.

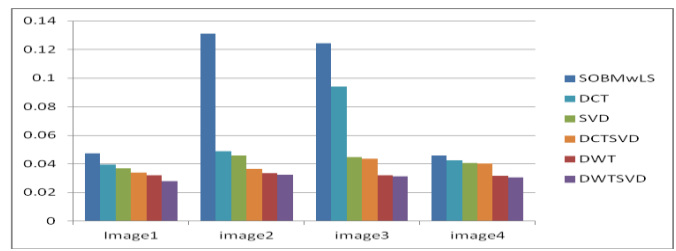

Figure.10: Comparison of Computation Times of Different Methods.

### 4.1 Detection Robustness against Intensity Variations

In the following Figure.11, a small part from the right bottom corner of the original image is copied and the intensities of the pixels are reduced by 50 before pasting to the bottom left corner of the image so as to obtain the test image. Now as these two regions have different intensity values for the pixels, the simple block matching methods (without DCT and PCA) detects no matching. But, as it can be seen from the Figure.11 (b) and (c) respectively, the DCT and PCA based method successfully identifies the duplicated regions. The duplicate regions as identified by the algorithm are shown in orange color.

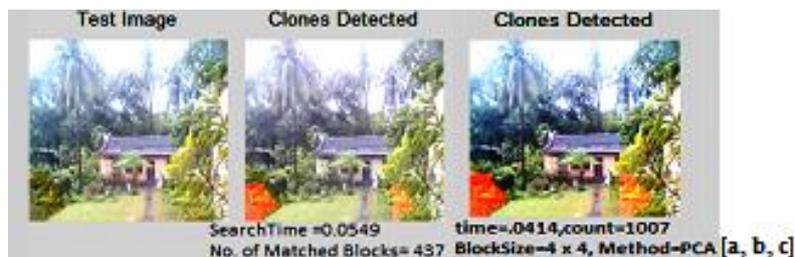

Figure.11: Detection of duplicate regions those differ in their intensities.

## 5. CONCLUSION AND FUTURE SCOPE

Cloning or copy-move forgery is one of the widely used image tampering methods. In this paper we have discussed different methods of cloning detection those successfully detect duplicated blocks in uncompressed images. We also have shown how the time complexity of the algorithms can be improved through DWT, SVD and how the DCT and PCA based methods can be effetely used to detect duplicated blocks even after brightness and contrast adjustments performed to the copy-pated regions. However, these methods fail to detect tampering in JPEG compressed images and unfortunately nowadays, almost all images are available in JPEG format. We are trying to extend our work to detect tampering in JPEG images as well.